\documentclass[prd,nofootinbib,preprintnumbers,amsmath,amssymb]{revtex4}

\usepackage{bm}
\usepackage{psfrag}
\usepackage{graphicx}

\newcommand{\be}{\begin{equation}}
\newcommand{\ee}{\end{equation}}
\newcommand{\bea}{\begin{eqnarray}}
\newcommand{\eea}{\end{eqnarray}}
\newcommand{\bml}{\begin{mathletters}}
\newcommand{\eml}{\end{mathletters}}

\begin{document}

\preprint{DCPT-04/15; hep-th/0405026}

\title{Smeared branes and the Gubser-Mitra conjecture}

\author{Paul Bostock}
\email{P.B.Bostock@durham.ac.uk}
\author{Simon F. Ross}
\email{S.F.Ross@durham.ac.uk}
\affiliation{Centre for Particle Theory, Department of Mathematical
  Sciences \\ 
University of Durham, South Road, Durham, DH1 3LE, UK  }

\date{\today}

\begin{abstract}
  We show that smeared brane solutions, where a charged black $p$-brane
  is smeared uniformly over one of the transverse directions, can have
  a Gregory-Laflamme type dynamical instability in the smeared
  direction even when the solution is locally thermodynamically
  stable. These thus provide counterexamples to the Gubser-Mitra
  conjecture, which links local dynamical and thermodynamic stability.
  The existence of a dynamical instability is demonstrated by
  exploiting an ansatz due to Harmark and Obers, which relates charged
  solutions to neutral ones.
\end{abstract}

\pacs{04.70.Bw,04.70.Dy,11.27.+d}
\maketitle

\section{\label{sec:intro}Introduction}

One of the most remarkable features of black holes is the connection
between properties of the classical solutions and thermodynamics. The
realisation that the laws of black hole mechanics (describing some 
dynamical aspects of black hole horizons) can be identified
with the laws of thermodynamics \cite{bar,bek} is at the origin of more than
30 years of work using black hole thermodynamics to gain clues about the
form of the quantum theory of gravity.

Black $p$-brane solutions in more than four spacetime dimensions are
richer dynamical systems, exhibiting new behaviours that have no
analogues for black hole solutions. In particular, some $p$-brane
solutions suffer from a classical instability discovered by Gregory
and Laflamme~\cite{GL1,GL2}. They showed that perturbations of the
metric with wavelength longer than some critical wavelength in the
extended directions grow exponentially. Thus, the perturbation breaks
translational invariance in the extended directions. The existence of
this instability raises two fundamental questions:
\begin{itemize}
\item
Given that a solution is unstable, what is the final state of the instability?
\item 
Can the connection between classical dynamics and thermodynamics be
extended to include this feature?
\end{itemize}
These questions have been the focus of considerable
activity. In~\cite{GL1}, a heuristic argument for the appearance of
the instability was advanced, based on comparing the
entropy of the $p$-brane to that of a periodic array of black
holes. This represents a first attempt to make a connection between this
instability and thermodynamics, and at the same time suggests that the
end state for the instability might be such an array of black holes. 

More recently, a more precise connection between dynamics and
thermodynamics was conjectured by Gubser and Mitra~\cite{GM}, who
suggested that \emph{a black brane with a non compact translational
  symmetry is classically stable if and only if it is locally
  thermodynamically stable}\footnote{Thermodynamic stability is taken
  to mean that the Hessian of the entropy (thought of as a function of
  extensive variables such as the charge and mass of the solution) has
  no positive eigenvalues.}. This conjecture was partially motivated
by the dual description of such black $p$-branes in string theory in
terms of a finite-temperature gauge theory on the $p$-brane's
worldvolume. A partial proof of this conjecture was given for a
certain class of $p$-branes by Reall~\cite{Reall}. The proof is based on
relating the threshold unstable mode, which has wavelength equal to
the critical wavelength, to a negative mode of the Euclidean black
hole solution. Further investigations of this relation were carried
out in~\cite{box,hirayama2,gubser:uni,kang}; the results so far
support the conjecture. Studies of the connection between dynamical
and thermodynamic instability which relax the requirement of
translational invariance appear in~\cite{hirayama,hubeny,hartnoll}.

Our aim is to extend these investigations of the conjecture to study
smeared branes: that is, we take a $p$-brane and smear it uniformly
over one of the transverse directions, and study stability to
perturbations in this smeared direction. This is a natural extension
of the investigation of $p$-branes
in~\cite{Reall,hirayama2,gubser:uni,kang}. The two classes of
solutions are related by T-duality, which implies that the
thermodynamics of the smeared branes is identical to that of the
$p$-brane with the same total number of extended directions. However,
the study of perturbations in the smeared direction is technically
more challenging; certain simplifications exploited in~\cite{Reall} no
longer apply. We will exploit recent advances in the construction of
{\it non-uniform} brane solutions to circumvent this problem. 

This work on non-uniform solutions was inspired by a contemporaneous
development concerning the second question above: Horowitz and
Maeda~\cite{HM} argued that the horizon could not pinch off, so the
end state of the instability could not be a collection of separate
black holes. Instead, they conjectured that the solution would settle
down to a non-translationally invariant solution with the same horizon
topology as the original $p$-brane.  Examples of solutions which are
non-uniform in one extended direction were subsequently found by
perturbing the neutral uniform black string by the threshold unstable
mode, which generates a branch of static non-uniform
solutions~\cite{Gubser,wiseman} (this was recently extended to higher
dimensions~\cite{sorkin}). These non-uniform black strings have too
large a mass to be the endpoint of the instability, but they show that
non-uniform solutions with regular event horizons do
exist\footnote{Preliminary numerical investigations of the endpoint of
  the Gregory-Laflamme instability were reported in~\cite{num}.}.

In~\cite{ansatz}, Harmark and Obers presented a useful ansatz for such
non-uniform solutions. Their ansatz, which will be reviewed in detail
in the next section, describes both vacuum black strings of the type
discussed in~\cite{Gubser,wiseman} and charged black branes smeared
over a transverse circle. The ansatz involves just two undetermined
functions, but in the vacuum case, it has been shown to be equivalent
to the general conformal ansatz, which involves three undetermined
functions~\cite{wiseman2,phase2}.

In this paper, we will use some of these results from the study of the
possible end states of the instability to show that smeared charged
black holes provide a counter-example to the Gubser-Mitra conjecture,
implying that the connection between dynamical instability and
thermodynamics is more complicated than previously thought. The key is
that in the Harmark and Obers ansatz, the full equations of motion are
satisfied if the two undetermined functions satisfy a system of
equations which are independent of the charge~\cite{ansatz}.  Hence,
any solution in the vacuum case gives a solution in the charged case.
In particular, the branch of non-uniform solutions meeting the black
string at the threshold unstable mode found in~\cite{wiseman} must
persist for non-zero charge. That is, there is a threshold unstable
mode for any charge. If we increase the charge sufficiently, the
uniform solution becomes thermodynamically stable; however, we take
the existence of the threshold unstable mode to indicate that it
remains dynamically unstable, violating the Gubser-Mitra
conjecture\footnote{In a recent paper~\cite{marolf}, it was suggested
  that gyrating strings might also give a counterexample to the
  Gubser-Mitra conjecture.}.

As in~\cite{Reall}, our analysis involves only the study of static
solutions, so we do not explicitly find the instability. However, the
fact that the `phase diagram' of static solutions is qualitatively
unchanged as the charge varies strongly argues that there is still an
instability. Our result certainly obstructs any attempt to extend the
argument of~\cite{Reall} to such smeared charged string solutions. 

It is surprising that the addition of a smeared charge, which
certainly affects the thermodynamic properties of the black brane
solution, does not affect the dynamics of the perturbations described
by the Harmark and Obers ansatz. It would be very interesting to
understand this observation from a dual field theory point of view.
Such an understanding might also help to see how the Gubser-Mitra
conjecture can be modified or reformulated in light of these results.

The next section reviews the essential features of the Harmark and
Obers ansatz, and uses them to show that there is a threshold unstable
mode for a smeared charged black hole. In the appendix, we discuss the
extent to which this ansatz provides a general description for charged
smeared branes. 

\section{Harmark and Obers' ansatz}
\label{sec:ansatz}

In~\cite{ansatz}, an ansatz for charged dilatonic black hole solutions
on a cylinder $\mathbb{R}^{d-1}\times S^1$ was introduced. This ansatz
was motivated by introducing a coordinate system which interpolates
between the usual black brane with transverse space $\mathbb{R}^{d}$,
which is a good description of a black hole on a cylinder of small
mass, and the black brane smeared on the transverse circle, which is a
good description at large mass. The ansatz is 
\bea
ds_{n}^{2}=H^{-\frac{d-2}{n-2}}\left(-fdt^{2}+\sum_{i=1}^{p}(dx^{i})^{2}+
  HR_{T}^{2}\left(f^{-1}AdR^{2}+\frac{A}{K^{d-2}}dv^{2}+KR^{2}
    d\Omega_{d-2}^{2}\right)\right)  
\label{ansatz metric}\\ 
e^{a\phi}=H^{2},\phantom{12} A_{01..p}= \coth \alpha
\left(1-H^{-1}\right),\phantom{12}
f=1-\frac{R_{0}^{d-3}}{R^{d-3}},\phantom{12}
H=1+\frac{R_{0}^{d-3}\sinh^{2}\alpha}{R^{d-3}},
\label{ansatz other}
\eea
where $A$ and $K$ are two unknown functions of $R$ and $v$ only, and
the total spacetime dimension $n = d+p+1$.  This solution has an event
horizon at $R=R_0$. 

The uniform smeared black $p$-brane is given by setting $A=K=1$. It's
thermodynamics are equivalent to those of the T-dual $p+1$-brane
solution. In particular, the mass and charge are\footnote{We take the
  $v$ and $x^i$ coordinates 
  to be periodically identified to allow us to write
finite expressions.} 
\be
M=\frac{\Omega_{d-2}2\pi R_{T} V_p}{16\pi
  G}(R_{T}R_{0})^{d-3}[(d-2)+ (d-3)\sinh^{2}\alpha],
\ee
\be \label{charge}
Q=\frac{\Omega_{d-2}2\pi R_{T}}{16\pi G}(R_{T}R_{0})^{d-3}(d-3)\sinh
\alpha \cosh \alpha, 
\ee    
while the entropy and temperature are
\be \label{st}
S = \frac{\Omega_{d-2}2\pi R_{T} V_p}{4G} (R_T R_0)^{d-2} \cosh \alpha,
\quad T
= \frac{d-3}{4\pi (R_T R_0) \cosh \alpha}. 
\ee
The statement of the conjecture uses the Hessian matrix of
derivatives of the entropy as the test for thermodynamic
stability. However, following~\cite{Reall,hirayama2,gubser:uni}, we
will assume that there is no charged field in the theory, so the
charge is not allowed to vary, and focus on the specific heat: we take
the condition for thermodynamic stability to be the positivity of the
specific heat, 
\be
C_Q = \left( \frac{\partial M}{\partial T} \right)_Q >0. 
\ee
It is easy to work out from these formulae that the
specific heat is negative at $Q=0$ for all values of $d$, but it becomes
positive above some critical charge if $d>5$. 

We now review the study of non-uniform solutions in this
ansatz. In~\cite{ansatz}, it was shown that when we impose the
equations of motion  
\bea
&R_{\mu\nu}-\frac{1}{2}\partial_{\mu}\phi \partial_{\nu}\phi
-\frac{1}{2}e^{a\phi}F_{\lambda\mu}F^{\lambda}_{\phantom{1}\nu}=g_{\mu\nu}\left(
  \frac{1}{4(2-n)}\right)e^{a\phi}F^{2}  
\label{gravity eom},\\ 
&\nabla^{2} \phi = \frac{a}{4}e^{a\phi}F^{2},\phantom{123}
\nabla_{\mu}\left(e^{a\phi}F^{\mu\nu}\right)=0,
\label{dilaton eom}
\eea 
the resulting system of equations for $A$ and $K$ is independent of
the charge (i.e., of $\alpha$), and hence also of the value of $p$
(since the extra dimensions $x^i$ decouple in the neutral case).
Furthermore, the boundary condition necessary to ensure regularity at
the horizon is simply that $A(R_0,v)$ and $K(R_0,v)$ are constants, so
the boundary conditions also do not involve the charge (this boundary
condition corresponds physically to requiring that the surface
gravity, and hence the temperature, is constant along the horizon).
This allows us to map the problem of finding a charged solution of the
form (\ref{ansatz metric}) to finding a solution in the uncharged
case.

One of the equations of motion can be solved algebraically for $A$ in
terms of $K$; this leaves a system of three second-order equations
which need to be satisfied by the function $K(R,v)$. Generically, such
a system is heavily over-determined; however, it was shown
in~\cite{ansatz} that the system is consistent to second order in
perturbation theory. This surprising result was elucidated
in~\cite{wiseman2,phase2}, where it was shown that in the neutral
case, the seemingly restricted ansatz taken above is in fact
equivalent to the most general ansatz consistent with the symmetries.
For the neutral black string, the Harmark and Obers ansatz reduces to
\be ds_{n}^{2}=-fdt^{2}+R_{T}^{2}\left(f^{-1}AdR^{2}
  +\frac{A}{K^{d-2}}dv^{2}+KR^{2}d\Omega_{d-2}^{2}\right)
\label{neutral ansatz}, 
\ee
where
\be
f=1-\frac{R^{d-3}_{0}}{R^{d-3}}.
\ee
Using staticity and spherical symmetry, the most general 
metric for a black string can be brought to the form 
\be
ds^{2}=-e^{2B}dt^{2}+e^{2C}\left(dr^{2}+dz^{2}\right)+e^{2D}d\Omega_{d-2}^{2},
\label{conformal}
\ee 
referred to as the conformal form, as the $(r,z)$ space is written
in conformally flat coordinates. Here $B,C$ and $D$ are functions of
$r$ and $z$ only. Since the latter form for the metric involves three
arbitrary functions, 
while the former only involves two, it seems like the former must be
more restrictive.  However, they are in fact equivalent if we further
assume the equations of motion are satisfied~\cite{wiseman2,phase2}. To
get from (\ref{conformal}) to (\ref{neutral ansatz}), we need to
choose $R(r,z)$ so that $e^{2B} = f$ is only a function of $R$, and
$v(r,z)$ so that there is no $dR dv$ cross term in the resulting
metric, and we get the appropriate relation between the $g_{RR}$,
$g_{vv}$ and sphere components of the metric. These conditions can in fact
be satisfied, subject to an integrability condition; in terms of the
conformal form of the metric, this condition is
\be
(\partial_{r}^{2}+\partial_{z}^{2})B+(\partial_{r}B)^{2}+(\partial_{z}B)^{2}+
(d-2)(\partial_{r}B\partial_{r}D+\partial_{z}B\partial_{z}D)=0.
\label{eom}
\ee
This condition is exactly
the $R_{tt}=0$ equation of motion for the three-function conformal metric
(\ref{conformal}), so if the equations of motion are satisfied, we can
pass from (\ref{conformal}) to (\ref{neutral ansatz}) by a coordinate
transformation. In appendix~\ref{cons}, we discuss the extent to
which this argument can be generalised to the charged case. 

The important point for our present purpose is that this implies that
any solution of the equations of motion describing a neutral black
string, uniform or non-uniform, can be written in the form
(\ref{neutral ansatz}). This provides a convenient framework for
discussing solutions. In~\cite{phase1}, Harmark and Obers constructed a
phase diagram summarising the known solutions in terms of two
parameters, the mass $M$ and a relative binding energy $n$ (which
provides a measure of the non-uniformity of the solutions). For
general charge, the mass is 
\be \label{genmass}
M = \omega (R_{T}R_{0})^{d-3}[(d-2)+ (d-3)\sinh^{2}\alpha]
\ee
and the binding energy parameter is 
\be \label{bind}
n = \frac{1-(d-2)(d-3)\chi}{(d-3)\sinh^{2}\alpha +(d-2)-(d-3)\chi},
\ee
where 
\be 
\omega=\frac{\Omega_{d-2}2\pi R_{T}}{16\pi G_{N}},
\ee
and $\chi$ parametrises the asymptotic falloff of the unknown
function $K$~\cite{ansatz}, and is hence independent of the
charge. 

\begin{figure}[!h]
\begin{center}
\psfrag{1/3}{$\frac{1}{3}$}
\psfrag{Mgl}{$M_{GL}$}
\psfrag{n}{$n$}
\psfrag{M}{$M$}
\psfrag{0}{0}
\includegraphics[scale=0.5]{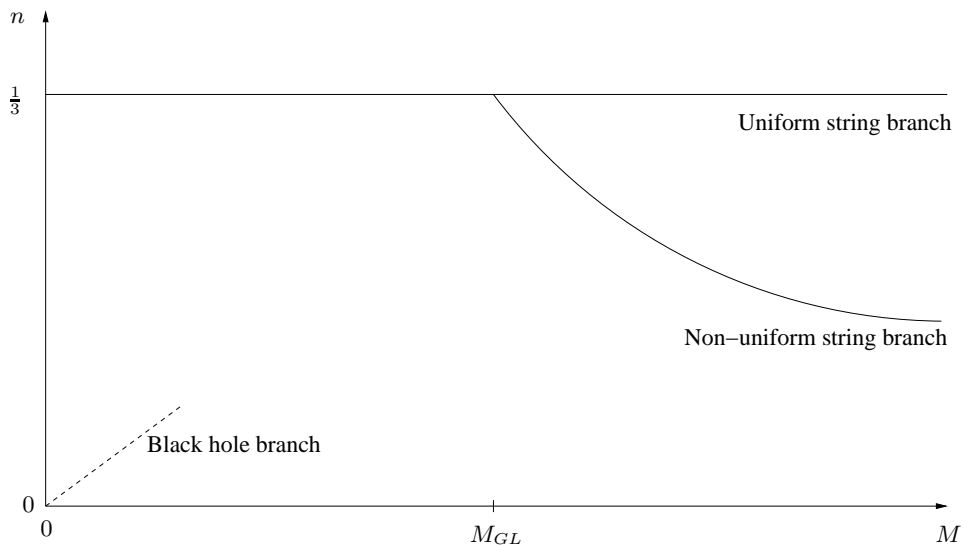}
\caption{\label{fig:1} Phase diagram for neutral solutions for a five-dimensional system on a circle.}
\end{center}

\end{figure}

The phase diagram for the neutral case given in~\cite{phase1} for a
five-dimensional system on a circle of radius $R_T$ is sketched in
figure~\ref{fig:1}.  

\begin{figure}[!h]
\begin{center}
\psfrag{1/3}{$\frac{1}{3}$}
\psfrag{Mgl}{$M_{GL}$}
\psfrag{n}{$n$}
\psfrag{M}{$M$}
\psfrag{0}{0}
\includegraphics[scale=0.5]{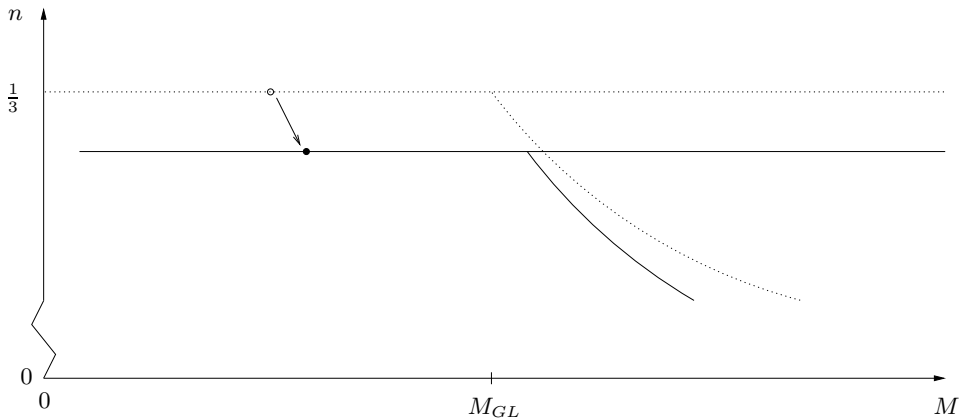}
\caption{\label{fig:2} The bold lines now refer to charged solutions.
  The diagram is qualitatively the same as in the neutral situation,
  shown as dotted lines for reference.} 
\end{center}

\end{figure}

Now since the equations of motion are independent of the charge, each
solution in figure~\ref{fig:1} gives a solution for every value of the
charge. Inspection of (\ref{genmass},\ref{bind}) shows that adding
charge increases the mass, as expected, and decreases $n$, enhancing
the binding effect.  Thus, if we plot $M$ vs $n$ in the charged case,
we get a qualitatively similar picture, as shown in figure~\ref{fig:2}.

We can see that on a circle of some fixed radius $R_T$, there is
always some critical value of the mass at which a non-uniform branch
joins on to the uniform smeared black hole branch. We can re-state
this in terms relevant for the Gubser-Mitra conjecture: for any given
value of the mass and charge, there is a finite wavelength at which
a threshold unstable mode occurs. We know that for zero charge, the
uniform black string is unstable to modes of longer wavelength.
Although we have not demonstrated the existence of the corresponding
dynamical instability explicitly in the charged case, the persistence
of the threshold unstable mode is strong evidence that it exists.

This result should be contrasted with the analysis of $p$-brane
solutions in~\cite{hirayama2,gubser:uni,kang}. In those studies, it
was found that for the ten-dimensional supergravity $p$-brane
solutions with $p \leq 4$, there is a threshold unstable mode for the
neutral case, but this mode goes off to infinite wavelength at a
critical value of the charge, signalling the disappearance of the
instability. This was found to occur at the same critical value of the
charge where the specific heat changes sign. 

In our case, by contrast, the threshold unstable mode exists all the
way up to extremality, even though the specific heat changes sign
before we reach extremality for cases with $d>5$. Thus, there are
smeared branes which are locally
thermodynamically stable, but possess a dynamical instability by the
argument of the previous section. This is a clear violation of the
Gubser-Mitra conjecture. Note that the wavelength of the threshold
unstable mode, which signals the onset of instability, is determined
by $R_0$, since the equations for $K$ are independent of charge. Hence
if we go near extremality by taking $R_0 \to 0$ and $\alpha \to
\infty$ keeping $M$ fixed, the wavelength of the unstable mode will go
to zero, suggesting that the instability will appear sufficiently
close to extremality for any compactified black string as well. 

\acknowledgments

We thank Ruth Gregory for discussions and collaboration at an early
stage of this project. We would also like to thank Troels Harmark,
Neils Obers and Mukund Rangamani for useful discussions. PB was
supported by a PPARC studentship. SFR was supported by the EPSRC.

\appendix

\section{Generality in the charged case}
\label{cons}

In section~\ref{sec:ansatz}, we reviewed the argument of
\cite{wiseman2,phase2} that the Harmark and Obers ansatz is consistent
in the neutral case. Since the equations for the unknown functions
$A(R,v)$ and $K(R,v)$ are independent of the charge, this also implies
that the ansatz is consistent in the charged case. It would still be
interesting, however, to ask if we can show that the {\it most general}
solution of the equations of motion with the assumed symmetries can be
written in the form (\ref{ansatz metric},\ref{ansatz other}) when we
include charge. 

We can easily show that the metric can be written in the form
(\ref{ansatz metric}) by an extension of the previous
argument. Starting from the 
3-function conformal form (\ref{conformal}), we can make the
redefinitions, 
\bea \label{redef}
e^{B} &\rightarrow& \bar{H}^{-\frac{d-2}{n-2}}e^{B},\\
e^{C} &\rightarrow& \bar{H}^{\frac{1}{n-2}}e^{B},\\
e^{D} &\rightarrow& \bar{H}^{\frac{1}{n-2}}e^{D},
\eea
for any function $\bar H$, so that (\ref{conformal}) becomes
\be
ds^{2}=\bar{H}^{-\frac{d-2}{n-2}}\left(-e^{2B}dt^{2}+\bar{H}e^{2C}\left(dr^{2}
    +dz^{2}\right)+\bar{H}e^{2D}d\Omega_{d-2}^{2}\right).    
\label{new conformal}
\ee 
If we now perform the same change of variables as was used
in~\cite{phase2} in the neutral case, 
\bea
R^{d-3}&=&R_{0}^{d-3}+\bar{r}^{d-3}\\
\hat{A}&=&f^{-1}AR_{T}^{2}\left(\frac{\bar{r}}{R}\right)^{2(d-4)}\\
\hat{K}^{d-2}&=&\frac{K^{d-2}}{f}\left(\frac{\bar{r}}{R}\right)^{2(d-4)}
\eea and then transform to the conformal form by making the further
transformation, \bea &\bar{r}=g(r,z),\phantom{123}
v=h(r,z)\\
&\partial_{r}g=e^{-(d-2)k}\partial_{z}h,\phantom{123}
\partial_{z}g=e^{-(d-2)k}\partial_{r}h.
\label{integrability}
\eea
We can bring the metric (\ref{ansatz metric}) in the
ansatz to the form (\ref{new conformal}) if
\be
g^{d-3}=\frac{R_{0}^{d-3}e^{2B}}{1-e^{2B}},
\ee
and
\bea
e^{2a}=\frac{e^{2c}}{(\partial_{r}g)^{2}+(\partial_{z}g)^{2}},\phantom{123}
e^{2k}=\frac{R_{T}^{2}}{R_{0}^2}e^{2D}e^{\frac{2(d-5)}{(d-2)(d-3)}B}
\left(1-e^{2B}\right)^{\frac{2}{d-3}}.
\label{match}
\eea
The system of equations in (\ref{integrability}) imply an
integrability condition which together with (\ref{match}) imply that
\be
(\partial_{r}^{2}+\partial_{z}^{2})B+(\partial_{r}B)^{2}+(\partial_{z}B)^{2}+
(d-2)(\partial_{r}B\partial_{r}D+\partial_{z}B\partial_{z}D)=0,
\label{cheom}
\ee 
the same integrability condition we had in the neutral case.

If we assume that the arbitrary function $\bar H$ introduced in the
redefinitions (\ref{redef}) is identified with the dilaton as in
(\ref{ansatz other}), i.e., $e^{a \phi} = \bar H^2$, we can show that this
integrability condition is again implied by the equations of motion.
The most general form for $F_{\mu\nu}$ consistent with the assumed
symmetries has only $F_{tz}$ and $F_{tr}$ non-zero; hence we can write
\be
F_{\lambda t}F^{\lambda}_{\phantom{1}t}=\frac{1}{2}g_{tt}F^{2},
\ee 
and so eliminating the dilation field using (\ref{dilaton eom}) from
the $t,t$ component of the graviton equation (\ref{gravity eom}) gives
us
\be
R_{tt}=g_{tt}\frac{n-3}{2(n-2)}\nabla^{2}(\ln \bar{H}).
\label{Rtt}
\ee 
This equation reduces to exactly (\ref{cheom}) for the metric given in
equation (\ref{new conformal}) and therefore we conclude that this
integrability condition is implied by the equations of motion. 

However, this is not yet enough to show that the general solution
takes the form (\ref{ansatz metric},\ref{ansatz other}): we have not
yet shown that $\bar H = H(R)$, and we have no coordinate freedom left
to redefine it. The problem can be simply stated in
coordinate-independent terms: in the charged case, there are two a
priori independent scalar quantities, namely the norm of the timelike
Killing vector $\partial_t$ and the dilaton. The ansatz (\ref{ansatz
  metric},\ref{ansatz other}) assumes a specific functional form for
both of these. While we can choose coordinates so that one of them
takes the specified form, it will not be possible to do this for both
of them in general, without using some additional information.

Thus, while it seems quite natural to us to assume that the ansatz
(\ref{ansatz metric},\ref{ansatz other}) describes the most general
solution of the equations of motion in the charged case as well, we
cannot show this by some analogue of the arguments
in~\cite{wiseman2,phase2}. Rather, verifying our belief would require
explicitly solving the equations of motion. We reiterate that this
question of generality is irrelevant to the argument in the body of
the paper, which required only the observation that uncharged
solutions of the form (\ref{ansatz metric},\ref{ansatz other}) lift to
charged solutions.


\end{document}